\documentclass{WileyMSP-template}
\usepackage{amssymb} 
\usepackage{color}
\usepackage[version=3]{mhchem} 
\usepackage{epstopdf} 

\usepackage{tabularx}
\usepackage{booktabs}
\newcommand{\tableheadline}[1]{\multicolumn{1}{c}{\small}}
\usepackage{caption}
\usepackage{subfig}  
\usepackage[T1]{fontenc}
\usepackage{wrapfig}
\usepackage{cite} 

\DeclareGraphicsExtensions{.pdf,.eps,.eps,.jpg,.mps}

\begin{document}

\title{Quantitative ultrafast electron-temperature dynamics in photo-excited Au nanoparticles}

\maketitle


\author{Maria Sygletou*}
\author{Stefania Benedetti}
\author{Marzia Ferrera}
\author{Gian Marco Pierantozzi}
\author{Riccardo Cucini}
\author{Giuseppe Della Valle}
\author{Pietro Carrara}
\author{Alessandro De Vita}
\author{Alessandro di Bona}
\author{Piero Torelli}
\author{Daniele Catone}
\author{Giancarlo Panaccione}
\author{Maurizio Canepa}
\author{Francesco Bisio}

\begin{affiliations}	
Maria Sygletou\\
OptMatLab, Dipartimento di Fisica, Universit\`{a} di Genova, via Dodecaneso 33, I-16146 Genova, Italy\\
Email: sygletou@fisica.unige.it\\
Stefania Benedetti\\
CNR-Istituto Nanoscienze, via Campi 213/a, 41125 Modena, Italy\\
Marzia Ferrera\\
OptMatLab, Dipartimento di Fisica, Universit\`{a} di Genova, via Dodecaneso 33, I-16146 Genova, Italy\\
Gian Marco Pierantozzi\\
Istituto Officina dei Materiali-CNR, Laboratorio TASC, Area Science Park, S.S. 14, Km 163.5, Trieste, I-34149, Italy\\
Riccardo Cucini\\
Istituto Officina dei Materiali-CNR, Laboratorio TASC, Area Science Park, S.S. 14, Km 163.5, Trieste, I-34149, Italy\\
Giuseppe Della Valle\\
Dipartimento di Fisica, IFN-CNR, Politecnico di Milano, Piazza Leonardo da Vinci 32,
I-20133 Milano, Italy\\
Pietro Carrara\\
Dipartimento di Fisica, Universit\`{a} di Milano, via Celoria 16, Milano, Italy\\
Alessandro De Vita\\
Dipartimento di Fisica, Universit\`{a} di Milano, via Celoria 16, Milano, Italy\\
Alessandro di Bona\\
CNR-Istituto Nanoscienze, via Campi 213/a, 41125 Modena, Italy\\
Piero Torelli\\
Istituto Officina dei Materiali-CNR, Laboratorio TASC, Area Science Park, S.S. 14, Km 163.5, Trieste, I-34149, Italy\\
Daniele Catone\\
Istituto di Struttura della Materia (ISM-CNR), Division of Ultrafast Processes in Materials (FLASHit), Area della Ricerca di
Roma Tor Vergata, Via del Fosso del Cavaliere, 100, I-00133 Rome, Italy\\
Giancarlo Panaccione\\
Istituto Officina dei Materiali-CNR, Laboratorio TASC, Area Science Park, S.S. 14, Km 163.5, Trieste, I-34149, Italy\\
Maurizio Canepa\\
OptMatLab, Dipartimento di Fisica, Universit\`{a} di Genova, via Dodecaneso 33, I-16146 Genova, Italy\\
Francesco Bisio\\
CNR-SPIN, C.so Perrone 24, I-16152 Genova, Italy\\
\end{affiliations}

	

\keywords{gold, plasmonics, ultrafast dynamics, photoemission, hot electrons}

\textcolor{red}{"This is the peer reviewed version of the following article: [Sygletou, M., Benedetti, S., Ferrera, M., Pierantozzi, G. M., Cucini, R., Della, G., Carrara, P., De, A., di, A., Torelli, P., Catone, D., Panaccione, G., Canepa, M., Bisio, F., Quantitative Ultrafast Electron-Temperature Dynamics in Photo-Excited Au Nanoparticles. Small 2021, 2100050], which has been published in final form at [https://doi.org/10.1002/smll.202100050]. This article may be used for non-commercial purposes in accordance with Wiley Terms and Conditions for Use of Self-Archived Versions."}

\newpage

\begin{abstract}
We measured the femtosecond evolution of the electronic temperature of laser-excited gold nanoparticles, by means of ultrafast time-resolved photoemission spectroscopy
induced by extreme-ultraviolet radiation pulses. 
The temperature of the electron gas was deduced by recording and fitting high-resolution photoemission spectra
around the Fermi edge of gold nanoparticles providing a direct, unambiguous picture of the ultrafast electron-gas dynamics. 
These results will be instrumental to the refinement of existing models of femtosecond processes in laterally-confined and bulk
condensed-matter systems, and for understanding more deeply the role of hot electrons in technological applications.\par

\end{abstract}

\newpage

\section{Introduction}

Following the impulsive photoexcitation of bulk materials and nanoparticles (NPs), a non-thermal electron population is generated. Such an out-of-equilibrium gas is characherized by a broad energy distribution that thermalizes on a sub-picosecond time scale via electron-electron ({\em e-e}) and electron-phonon ({\em e-p}) collisions, giving rise to a high-temperature electron gas. The hot gas eventually releases its excess energy to the ion lattice through electron-phonon ({\em e-ph}) scattering on the picosecond ({\em ps}) time scale, and the lattice heat is then dissipated to the environment via phonon-phonon ({\em ph-ph}) interactions within few hundred \textit{ps} \cite{abinitio2016,Bisio2017,Saavedra2016,Yonatan2019,Ferrera_ACSphotonics_2020}. \par

These dynamic processes and the energy redistribution of the charges following the photoexcitation \cite{Kumar2016,Saavedra2016,Brongersma2015,Baffou2011}
lie at the very heart of some of the most interesting light-induced physical phenomena, such as photocatalysis \cite{Guo2013,bian2013,Clavero2014b,kale2014} and solar energy conversion \cite{Ferry2010,Atwater2010,kovalenko2010,Stratakis2013,Sygletou2017}. In particular, the ultrafast relaxation of the high-energy nonthermal electrons plays a pivotal role in determining the final outcome \cite{liu2018,smith2018}.

In metallic NP systems, the photoexcitation processes are made more interesting by the possible involvement of strong collective electron-gas excitations known as localized surface plasmons (LSPs), that promote strong light-matter interaction at the nanoscale \cite{Kravets2018}.
The ultrafast decay pathway of LSPs has attracted a lot of scientific interest due to the spectral tunability of the resonance \cite{SEPULVEDA2009}, its exploitabiliy in light harvesting\cite{Stratakis2013,Atwater2010} and its key role in hot charge-carrier generation \cite{Manjavacas2014}.  

Variables such as the characteristics of exciting radiation (fluence, wavelength, pulse duration, {\em etc.}) as well as the 
constituent material, the size and shape of the NPs, can all play a crucial role in the {\em e}-gas heating and in the energy-relaxation pathway \cite{Taylor2014,Herrmann2014,Gonzalez2017,Catone2018,Christofilos2003,loukakos2004,Minutella2017,Hartland2017,Zhang2018,gallardo2019}. 
Additionally, the temperature dependence of the material paratemers (e.g.~electronic specific heat, \cite{Stoll2014,Magnozzi2019,Ferrera2019} interface thermal conductivities \cite{Sayed2001}, {\em etc.}) will all affect the actual relaxation dynamics. 
 
The current understanding of ultrafast relaxation processes in plasmonic nanosystems rests upon ultrafast time-resolved optical and,
to a lesser extent, electronic spectroscopies, \cite{Feldstein1997,Hartland1998,loukakos2004,Wang2015,Harutyunyan2015} that
however mostly yield {\em indirect} information about  the time-dependent electron-gas or ion-lattice temperature \cite{Broek2011,Plech2017}.
Theoretical models, on the other hand, are becoming more and more refined, yet cannot handle,  so far, the complexity of
real systems \cite{Richardson2009,kovalenko2010,Manjavacas2014,Saavedra2016,Petek2021}.

In this work, we report the measurement of the ultrafast electron-temperature ($\vartheta_e$) dynamics within an ensemble of plasmonic  gold NPs,
laid onto a transparent conductive oxide support, thereby allowing to perform photoemission experiments
while allowing a degree of electron confinement sufficient to preserve the LSP resonance.
Experiments were performed in pump-probe configuration by photo-exciting the NPs close to the LSP resonance and collecting ultrafast time-resolved photoemission spectra (tr-PES). 

$\vartheta_e$ was assessed exploiting its own definition expressed by the Fermi distribution function, without
resorting to any kind of model. 
By executing ultrafast pump-probe measurements on Au NPs within the Fermi-edge energy window, the $\vartheta_e$ evolution as a function of 
the time delay elapsed since the exciting pulse could therefore be retrieved. 
We observed a fast evolution of the electronic temperature within the first {\em ps} after excitation, detecting a $\vartheta_e$ peak at 780-840 {\em fs} delay, followed by its gradual
relaxation towards environment temperature.
Whereas the general trend is in agreement with the current understanding of ultrafast relaxation dynamics,
the emerging of quantitative discrepancies with theoretical predictions underscore the key role of {\em direct} ultrafast measurement of the electronic temperature in order to correctly evaluate the ultrafast transient response of nanosystems. In this respect our quantitative, model-free measurement of the electron-temperature dynamics represents a major advancement for
significantly improving the future understanding and modelling of ultrafast dynamic processes in real-world nanoscale systems.




\section{Results}

\begin{figure}
		\centering
		\includegraphics[width=0.6\linewidth]{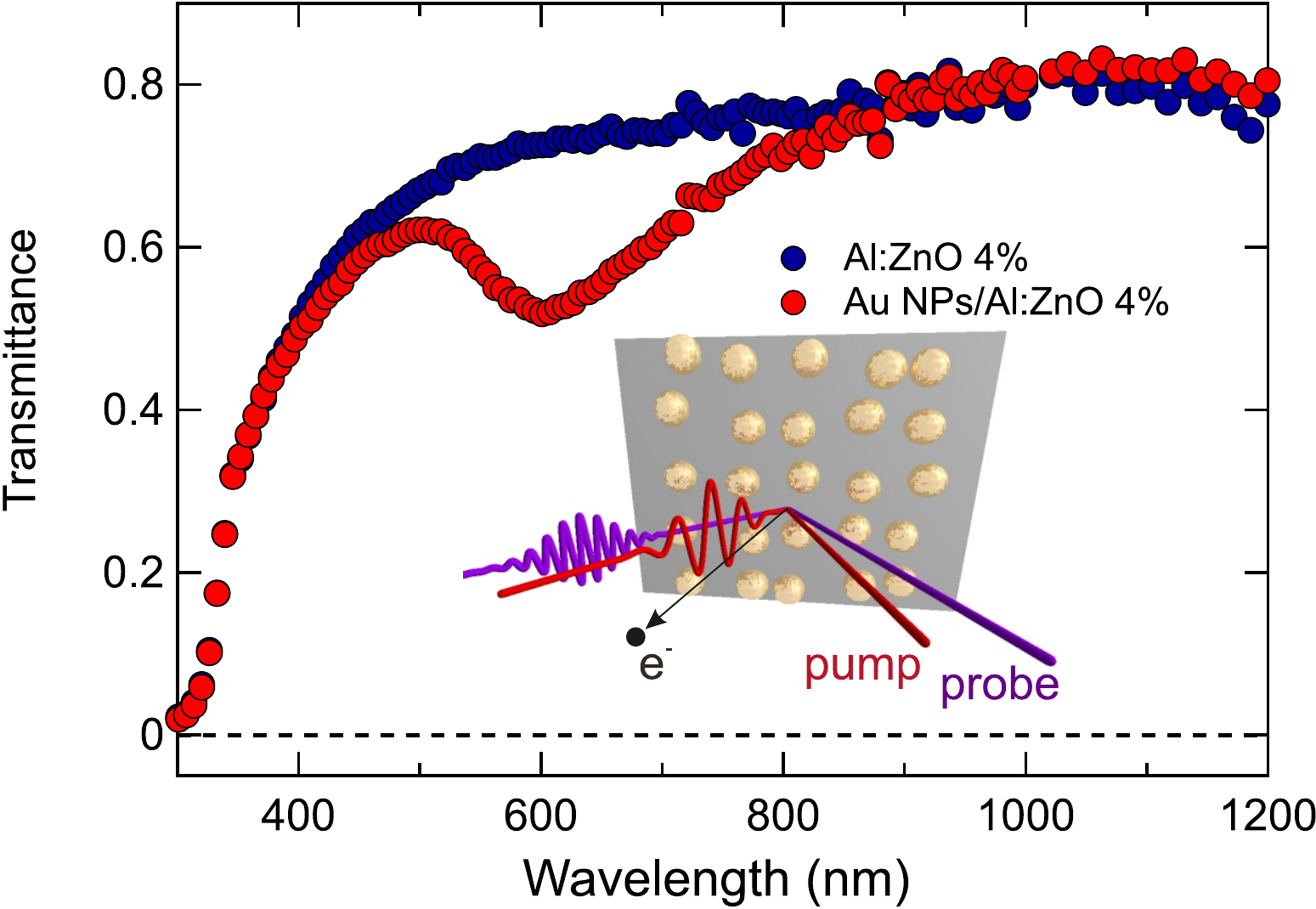}
		\caption{Transmission spectra of 4 at.\% AZO film on MgO substrate with (red markers) and without (blue markers) Au NPs.
Inset: schematic diagram of the experimental setup.}
		\label{trans_abs}
\end{figure}

In Figure \ref{trans_abs} we report the transmission spectra of the bare substrate, consisting of a 150 nm thick Al-doped ZnO (AZO) film deposited on a MgO substrate (blue markers)
and of Au NPs deposited onto it (red markers).
The broad transmission dip around 600 nm in the red spectrum is the fingerprint of the LSP resonance.
This was confirmed by comparing the experimental results of Figure \ref{trans_abs} with electromagnetic simulations in the quasi-static limit,
performed according to the same model reported in Ref.\cite{Ferrera_ACSphotonics_2020}, that has been carefully validated on both stationary and transient broadband spectroscopy measurements of the same samples (see the SI). 

The samples were measured by tr-PES in pump-probe configuration at the SPRINT Lab @ CNR-IOM \cite{Cucini2020}. The pump was an ultrashort pulse with wavelength $\lambda= 650$ nm while the probe was an extreme-ultraviolet (EUV) pulse obtained by high-harmonic generation (HHG, photon energy 16.9 eV).

\begin{figure}
\centering
		\includegraphics[width=0.6\linewidth]{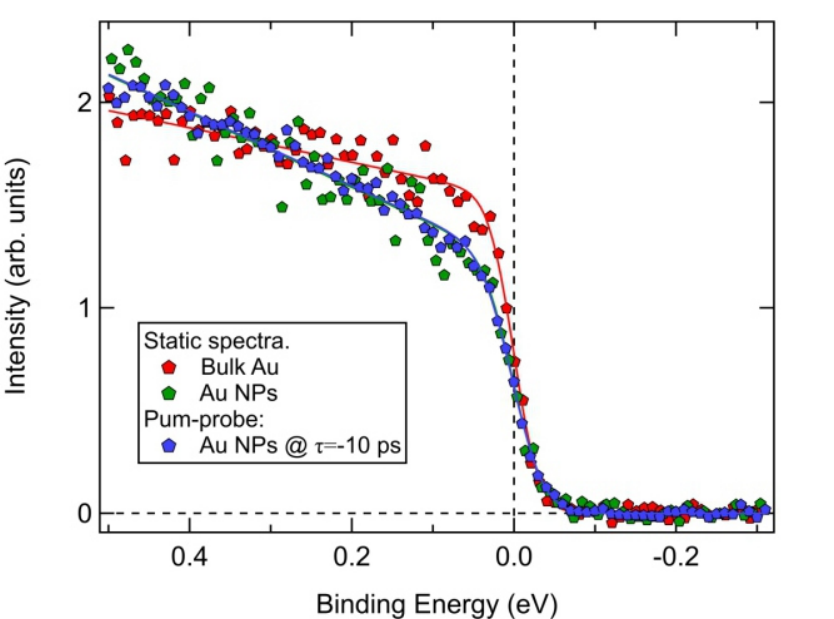}
		\caption{Experimental PE spectra in the Fermi edge region. Stationary data from bulk Au (red markers) and Au NPs (green markers) are reported. 
Pump-probe spectra of Au NPs at negative delay ($\tau$=$-$10 {\em ps}) are shown as blue open markers. 
The respective fitting curves according to equation (\ref{FermiDirac}) (solid lines) are reported.}
\label{static_pp_FE_fit}
\end{figure}

In Figure \ref{static_pp_FE_fit} we report three PE spectra measured in the Fermi-edge region of Au as a function of the binding energy (BE) $E_F-E$, $E_F$ being the Fermi energy and $E$ the electron energy.
The red symbols represent a reference measurement on a thick Au foil, performed with the EUV radiation only (blocking the pump beam).
We refer to these measurements as {\em stationary} spectra.
The green markers represent the stationary spectrum of Au-NPs/AZO/MgO, and the blue open markers the 
corresponding pump-probe spectrum, at delay time $\tau =-10$ ps, where no pump-induced effect are present.
The delay time $\tau$ is defined as the time interval between the excitation of the pump and the arrival of the probe pulse.
The continuous lines are the best fits to the experimental data performed by means of 
the function 
\begin{equation}
F(E)=\left[\ell(E)\cdot\left(\frac{1}{e^{\frac{\left(E-E_{F}\right)}{k_{B}\vartheta_e}+1}}\right)\right]*g(E,w)
\label{FermiDirac}
\end{equation}
where $\ell(E)$ is a linear function of energy accounting for the linear slope of the spectra below $E_F$ and for the dark-count contribution, $k_{B}$ is is the Boltzmann constant, $g(E,w)$ is a Gaussian function characterized by its width $w$ accounting for the finite experimental resolution, and $*$ denotes the convolution operator.
The best fit to the bulk-Au spectrum of Figure \ref{static_pp_FE_fit} was obtained
by fixing the experimental resolution at the instrumental value of $w=35$ meV 
and fitting $\ell(E)$ and $\vartheta_e$;
a temperature value of $\vartheta_e=(137 \pm 17)$ K was found.\\
The two NP spectra (blue and green markers in Figure \ref{static_pp_FE_fit}) were measured at the same sample temperature, yet
exhibit a slight smearing of the Fermi edge accompanied by a 
small spectral shift to higher binding energies with respect to the reference bulk Au.
Both phenomena are due to well-known finite-state effects in photoemission, typical of spatially-confined systems
\cite{smearing2004,Kurtz2007}, and can be {\em effectively} represented by a slight decrease of the energy resolution, and allowing for a small shift of $E_F$ in the fitting function.
The best fits of the Au-NP spectra were then performed holding $\vartheta_e$ at $137$ K
while allowing $\ell(E)$, $E_F$ and the experimental resolution to vary.
The best agreement was found for  $E_F$ downshifted by 1.5 meV with respect to the bulk-Au spectrum,
and the  experimental resolution $w=52\pm12$ meV.
The value thus found for $E_F$ was set as the zero of the binding energy for all subsequent NP spectra.
We point out that, strictly speaking, finite-state effects in photoemission distort the pure Fermi-Dirac distribution. However, NP spectra are still very nicely fitted using Eq. \ref{FermiDirac}, meaning that the effective representation of the spectra by a FD distribution with decreased energy resolution suits the purpose of extracting an electronic temperature from our spectra.

The pump-probe spectrum at negative delays (blue markers) and its stationary counterpart seem basically overlapping.
Running a fitting procedure, the instrumental broadening is the only parameter that looks changed, from $w=52\pm12$ meV to $w=56\pm4$ meV, a difference that
is comparable with the experimental uncertainty.
For the analysis of the time dependent data, the experimental broadening was kept fixed at $w=56$ meV,
so that $\vartheta_e$ is the only parameter left accounting for the width of the distribution around $E_F$.

\begin{figure}
		\centering
		\includegraphics[width=0.6\linewidth]{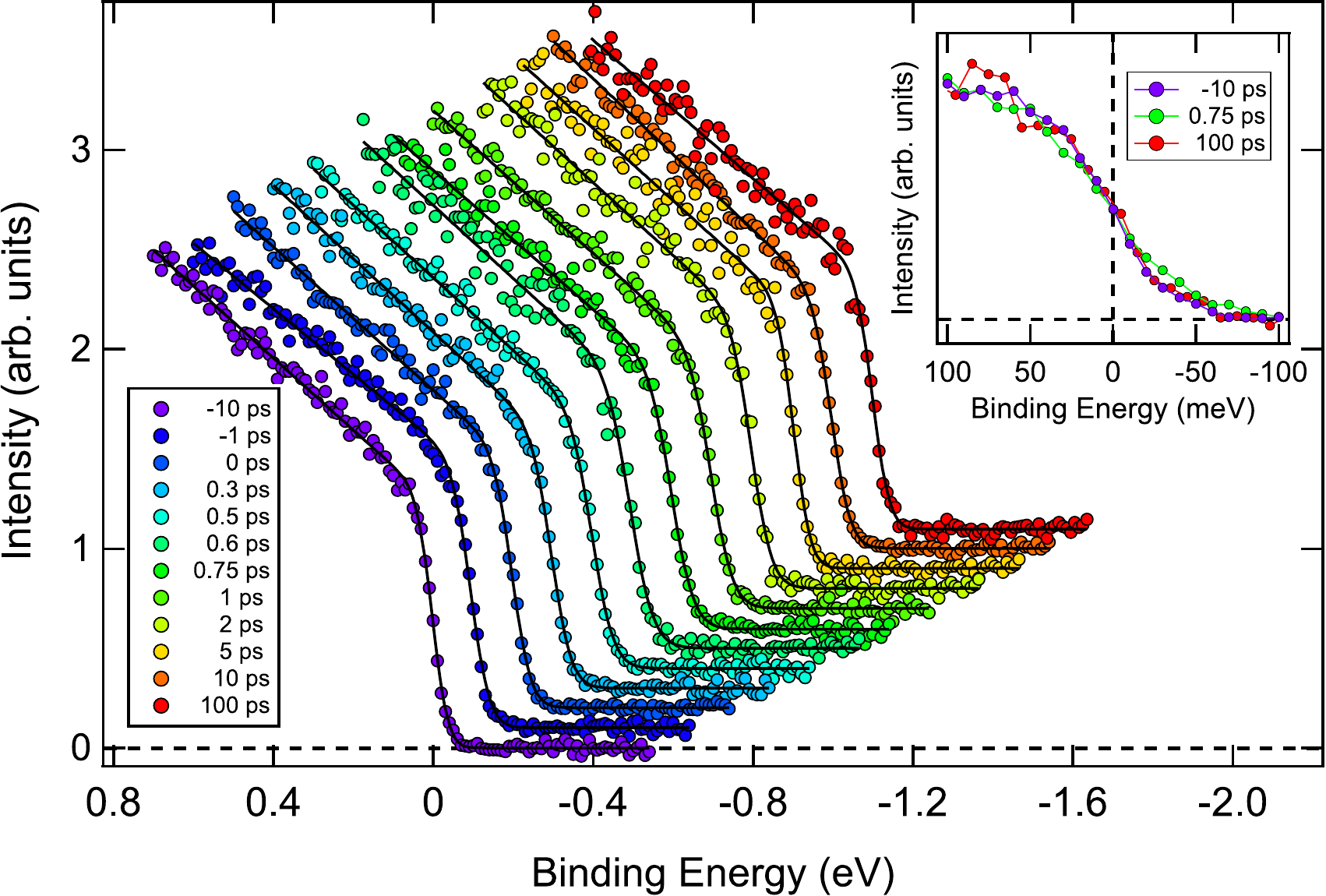}
				\caption{Time-resolved pump-probe spectra of Au NPs at the Fermi edge (markers). Best fits according to equation (\ref{FermiDirac}) (solid  lines). The curves have been offset in energy and intensity for the sake of clarity.
The inset shows spectra acuired in correspondence of three representative delay values (-10 ps, 0.75 ps and 100 ps respectively).}
		\label{FE_pp_double}
	\end{figure}

In Figure \ref{FE_pp_double} the pump-probe spectra of the FE region of Au NPs at various time delays (from -$10$ ps to $100$ ps), are reported. 
The curves have been offset in energy and intensity for the sake of clarity.
In the inset, a Fermi-energy zoom of PE spectra acquired in correspondence of 3 representative delay values  ($\tau=-10$ ps, $\tau=0.75$ ps and $\tau=100$ ps respectively) is reported. 
At a first glance, only very tiny variations are seen for the different delays (see inset), therefore in order to extract the experimental $\vartheta_e(\tau)$ curves, the application of a fitting procedure is deemed necessary.
In order to improve the reliability of the $\vartheta_e$ estimation, two different independent methods were applied.\\ 
In the first method, each curve in Figure \ref{FE_pp_double} (solid lines) was fitted according to equation (\ref{FermiDirac}), leaving
$\ell(E)$, $E_F$ and $\vartheta_e$ as free parameters and fixing the effective experimental resolution at $w=56$ meV. 
Although $E_F$ was a free parameter, no shift of the relative position of the Fermi edge was observed for the various delay-dependent spectra. 
The values of $\vartheta_e(\tau)$ so obtained are reported in Figure \ref{2methods_double}
as the red markers.\\
In the latter method, we exploited the mathematical relationship between $\vartheta_e$ and the first derivative of the FD distribution at $E=E_F$:

\begin{equation}
\label{LF}
{\vartheta_e}=-\left.{\frac{1}{4k_{B}}}{ \left(\frac{\partial f(E)}{\partial E} \right)^{-1}}\right|_{E=E_{F}},
\end{equation} 

and performed linear fits of the experimental spectra around $E_F$. 
We applied a mathematical procedure to correct the 
$\vartheta_e$ overestimation due to the finite experimental resolution and
we obtained the $\vartheta_e (\tau)$ curve reported in Figure \ref{2methods_double} as the blue markers.
Details of the fitting procedure can be found in the SI.

The values of $\vartheta_e$ in Figure \ref{2methods_double}, extracted by the two fitting methods follow the same general trend,
with some discrepancy that will be discussed later.
From $\tau=0$ to $\tau=600-750$~fs (depending on the fitting method), we observe an increase of the parameter $\vartheta_e$, peaking at $286\pm29$ K ($232\pm21$ K) depending which fitting method is employed.
Following this maximum, $\vartheta_e$ shows a decreasing trend.
The $\vartheta_e$ values deduced by the FD fitting of the tr-PE spectra are  larger than the ones obtained from the slope method (in some cases quite remarkably, like {\em e.g.} for $\tau=-1$~ps), and
some of the data points appear scattered with respect to an idealized smooth trend;
this is due to the extreme sensitivity of the analysis procedure to very fine details
of the tr-PE spectra within a very narrow energy window around $E_F$.
Such a sensitivity implies that stochastic noise at specific data points within this narrow energy window can affect the experimental $\vartheta_e$
even though the overall signal-to-noise ratio of the spectra is remarkably good, and also explains slight differences between the two analysis methods.

\newpage

\section{Discussion}

As anticipated in the introduction, the energy absorbed from the pump pulse generates, via ultrafast dephasing of the collective surface plasmon excitation~\cite{theoretical2014,theory2015,liu2018,ACSNANO2020}, two types of excited carriers (see e.g.~\cite{Govorov2013,Govorov2014,Brongersma2015,Chang_ACSEL_2019} for an overview): highly energetic {\it nonthermal carriers} and {\it thermalized hot carriers}, described in terms of the electronic temperature $\vartheta_e$. Concerning the nonthermal carriers, recent experiments report that the actual distribution of non-thermal carriers in a plasmonic metal \cite{Petek2021} may differ from the calculated ones in e.g. \cite{theoretical2014} and \cite{theory2015}. In particular, it is shown that under excitation with photon energies matching the bulk plasmon energy, the bulk plasmon decay preferentially excites photoelectrons from the Fermi level rather than producing a broad energy distribution of hot electrons and holes, a process referred to “plasmonic photoemission” \cite{Petek2021}. In our setup, with pump photon energy much lower than bulk plasmon energy, this phenomenon can’t be observed.
Despite of the controversies on the precise energy distribution and lifetime of nonthermal carriers, a general consensus exists regarding two key issues:
(i) the nonthermal carriers are distributed on a broad range of energies, comprising electronic states far from the Fermi level, contrary to thermalized carriers whose occupation probability is perturbed with respect to the equilibrium temperature FD distribution only close to the Fermi level; (ii) the dynamics of $\vartheta_e$ reflects the dynamics of energy release from the nonthermal electronic population to the thermal one. Therefore, as long as nonthermal carriers persist, the electronic temperature $\vartheta_e$ can't describe the whole electronic subsystem~\cite{Saavedra2016}. However, under the condition that the relative weight of the non-thermal population with respect to the thermalized {\em e}-gas is negligible, one can extract from the Fermi-edge spectra the effective temperature of the thermalized fraction of the electron gas. This is precisely the case of our experiments, given the low fluence of the pump pulses. In this respect, the increase of $\vartheta_e$ that we observed in the first few hundred \textit{fs} is due to the gradual heating of the thermalized fraction of the electron gas, due to the relaxation  of the non-thermal carriers \cite{tzortzakis2000,Christofilos2003,Minutella2017,hartland2004,groeneveld1992,sun1994,Imamura2006}. 
The maximum value of $\theta_e$ corresponds to the completion of the {\em e}-gas thermalization, while
its subsequent decrease is ascribed to the gradual thermalization of the electron gas with the lattice.

Concerning the nonthermal carriers, recent experiments report that the actual distribution of non-thermal carriers in a plasmonic metal \cite{Petek2021} may differ from the calculated ones in e.g. \cite{theoretical2014} and \cite{theory2015}. In particular, it is shown that under excitation with photon energies matching the bulk plasmon energy, the bulk plasmon decay preferentially excites photoelectrons from the Fermi level rather than producing a broad energy distribution of hot electrons and holes, a process referred to “plasmonic photoemission” \cite{Petek2021}. In our setup, with pump photon energy much lower than bulk plasmon energy, this phenomenon can’t be observed.

\begin{figure}
		\centering
		\includegraphics[width=1.0\linewidth]{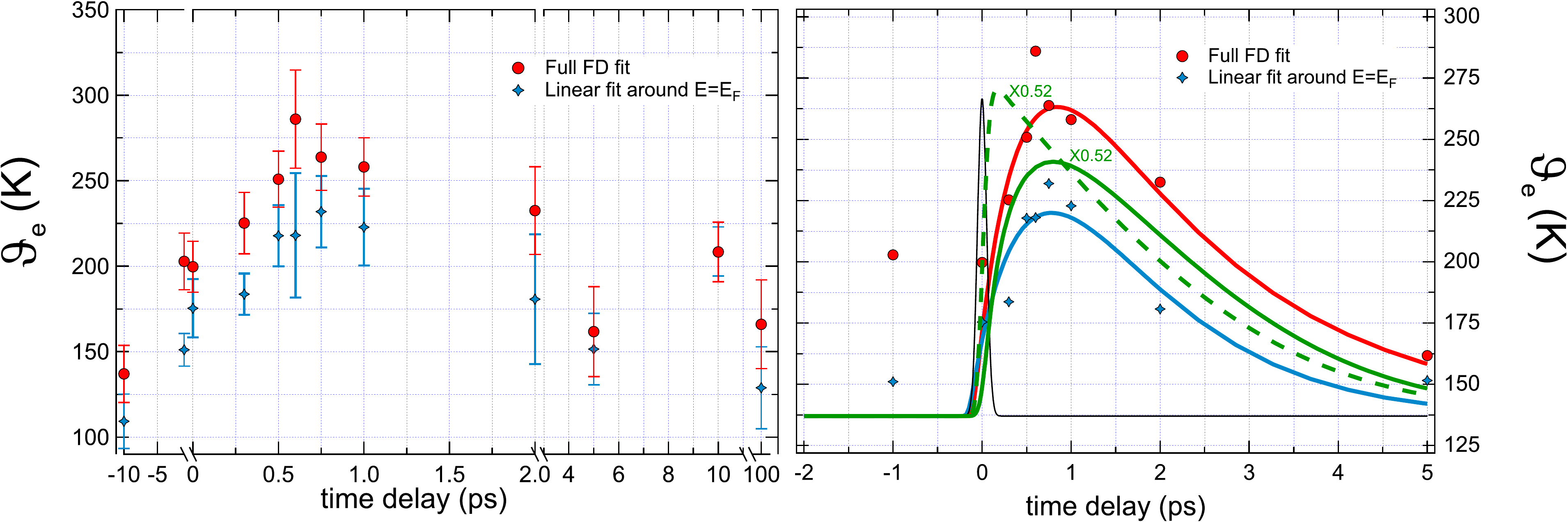}
		\caption{Left panel: $\vartheta_e(\tau)$ values extracted from the experimental spectra of Figure \ref{FE_pp_double}. Blue and red markers correspond to values extracted according to two different analysis methods (see text for details). Right panel: blue and red markers: experimental $\vartheta_e(\tau)$ as in the left panel. Blue and red solid lines: fits of the experimental data by means of the sum of an exponential rise function, an exponential decay function and a step function convolved with a Gaussian simulating the instrumental response function (IRF) of the system (see text for details).
Green lines: simulated $\vartheta_e(\tau)$ according to the three-temperature model (3TM) with $\tau_{ee} = 410$~fs (solid trace) and $\tau_{ee} = 30$~fs (dotted trace). The simulated $\vartheta_e(\tau)$ is rescaled by a factor of $0.52$ for better comparison with the measurements in terms of temporal dynamics. Black line: the Gaussian profile of the driving term in the 3TM (FWHM equal to 120 fs).}
		\label{2methods_double}
	\end{figure}

In order to validate our analysis, our experimental data must be tested against appropriate theoretical estimations.
Indeed, other spurious or unaccounted effects will not give rise to similar $\vartheta_e$ dynamics, neither in terms of the intensity of the effects
nor in terms of temporal dynamics.\\
Firstly, we performed an elementary calculation for estimating the order of magnitude of the maximum temperature increase of the electron gas, $\Delta \vartheta_e^{max}$. 
Under the assumption that that the electromagnetic energy absorbed by the
NP is used to homogeneously heat the electron gas, that the dielectric function
of Au is constant over the duration of the excitation pulse, and that no heat dissipation
to the environment has occurred (an approximation valid within few ps after the
exciting pulse) $\Delta \vartheta_e^{max}$ can be calculated as \cite{Baffou2011}:

\begin{equation}
\label{cal_T} 
\Delta \vartheta_e^{max}=\frac{F\cdot \alpha}{t\cdot\rho_{e}}\int_{pulse}\frac{dT}{C_{e}(T)}
\end{equation}

where $t$ is the effective thickness of Au, $\rho_{e}$ and $C_{e}(T)$ are the electron density and electron heat capacity of bulk Au, respectively, $F$ is the fluence and $\alpha$ is the absorption coefficient at $\lambda_{pump} = 650$~nm (defined as the fraction of impinging EM energy absorbed by the NPs).  


The experimental fluence, assuming the probe beam perfectly centered onto the, much larger, pump beam is $F= 0.19\pm0.03$ $\mathrm{J/m^2}$.
The absorption coefficient, deduced from the data in Figure \ref{trans_abs}-top at $\lambda = \lambda_{pump}$, assuming no radiation scattering from the NPs, was $\alpha = T_{substr.} -T = 0.18$, while  $t = 3$~nm. 
From the above we obtain $\Delta \vartheta_e^{max} = 487\pm10$~K, a value  higher than the experimental values by a factor of 3 to 4, depending on the fitting method. 
Given the simplifictions involved, this can be considered a fair agreement, yet
a more solid validation is provided by resorting to a dynamical model for the hot-electrons, where the latter simplifications are lifted. 
For this purpose we adopted the so-called Three-Temperature Model (3TM), originally introduced by Sun and coworkers for thin gold films,\cite{Sun_PRB_1994} and then  exploited and validated on optical pump-probe experiments in plasmonic~\cite{Zavelani_ACSP_2015, Della_Valle_JPCL_2013} 
and even all-dielectric~\cite{Gaspari_NL_2017} nanostructures. 
In short, the model accounts for the dynamics of three energy degrees of freedom in photoexcited metallic structures:  
the excess energy density stored in the population of nonthermal electrons ((directly coupled to the pump pulse), $N$, the temperature of the hot thermalized charges, $\vartheta_e$, and the lattice temperature, $\vartheta_l$. The coupling between $N$ and $\vartheta_e$ is dictated by an electron-electron thermalization rate, given in terms of the inverse of the {\it electron-electron thermalization time constant,} $\tau_{ee}$. The electron and lattice temperatures are coupled to each other by the so-called {\it electron-phonon coupling coefficient,} $G$. Further details on the 3TM and related parameters are provided in the Experimental Section.

The 3TM is here exploited to address a clearcut comparison with atomistic calculations that have recently pointed out key novelties on the ultrafast relaxation of nonthermal and thermal electrons.
The evaluation of the rise time of the experimental temperature dynamics was obtained by fitting the data with the sum of an exponential rise function, an exponential decay function and a step function convolved with a Gaussian that simulates the instrumental response function (IRF) of the system (Eq. 2 of the Supporting Information (SI)). In Figure \ref{2methods_double}-right the solid lines represent the curves resulted from these evaluations for the two fitting methods (red line and blue line). The rise times exctracted with this fitting procedure are 600 fs and 850 fs for data obtained with the first and second method, respectively.
The electron temperature dynamics for the thermalized fraction of the electron gas retrieved from 3TM simulations is also shown in Figure~\ref{2methods_double}-right (solid green line). To this aim, in the simulations we assumed $\tau_{ee}$ and $G$ as free parameters to be determined by quantitative comparison with the dynamics of the measured $\vartheta_e$.
The values of $\tau_{ee}$ and $G$ obtained from the simulations, according to our experimental data, turned out to be  $410$~fs and $0.91 \times 10^{16}$~W~ m$^{-3}$~K$^{-1}$, respectively. 
The theoretical prediction from the 3TM is in good agreement with the measured dynamics of $\vartheta_e$, yet the simulation apparently overestimates the temperature rise by a factor 2. Such a discrepancy is most likely ascribable to the overestimation of the energy absorbed by the NPs.
This is plausible since the modeling of the plasmonic response of the NPs neglects collective EM-coupling effects in the NP ensemble,
replacing them by an effective increase of the dielectric function of the NP environment. Also, the inhomogeneous broadening effects in the sample, arising from the dispersion of size inherent in the self-organized fabrication process, was mimicked by increasing the Drude damping of gold. Similar to what observed in previous studies~\cite{Della_Valle_PRB_2015, Silva_PRB_2018}, this approach enables one to correctly reproduce the broadening of the extinction spectrum, but at the expense of an overestimation of the absorption contribution by approximately a factor of 2. Moreover, in the low temperature regime of our experiments, the imaginary part of gold permittivity is lower than at room temperature~\cite{Bouillard_NL_2012}, which is another possible reason why our calculations overestimates absorption. Taken together, these aspects highlight the importance of performing a {\em direct} measurement of the $\vartheta_e$ dynamics, in order to have available a solid reference for theory and modelling.
Anyway, the results retrieved by the 3TM can be exploited to rationalize the outcome of the experimental measurements in terms of the temporal dynamics of $\vartheta_e$, which in the perturbative regime of our excitation will be not affected by the total amount of energy released to the nanostructures.

First, note that in order to match the long rise time of the hot-electron temperature observed in the experiments (with temperature peak achieved at $780-840$~fs) the $\tau_{ee}$ value in the 3TM turned out to be about two-times larger than that reported in previous papers (see e.g.~Refs.~\cite{Sun_PRB_1994, Zavelani_ACSP_2015}). However, the new estimate is in agreement with recent results from atomistic calculations. Actually, Govorov's and coworkers have shown that the nascent distribution of the out-of-equilibrium electrons ought to comprise more prominent contributions close to the Fermi level compared to the simpler assumption of a homogeneous energy distribution made within the more classic 3TM (see Fig.~2 in Ref.~\cite{Chang_ACSEL_2019}). This result, combined with the prediction that electronic states close to the Fermi level results in lower electron-electron scattering rates (see Fig.~3 in Ref.~\cite{Bernardi_NatComm_2015}), points in favor of a larger value for $\tau_{ee}$ compared to that commonly assumed in the literature. Moreover, note that the attempt to reduce $\tau_{ee}$ to the inverse of the electron-electron scattering rate, corresponding to about 30~fs~\cite{Bernardi_NatComm_2015}, dramatically fails to reproduce the correct dynamics of $\theta_e$ (dotted green trace in Figure \ref{2methods_double}-right).
This is in line with what recently reported by Nordlander and coworkers~\cite{liu2018}, also in collaboration with one of the present authors~\cite{Schirato_NatComm_2020}. Actually, even though individual scattering events occur on a time scale of few femtoseconds the full thermalization requires hundreds of these events and the overall relaxation time of the nonthermal carriers was thus estimated to be of the order of several hundreds of femtoseconds (see Fig.~5 Ref.~\cite{liu2018})).

Second, note that the decay of the measured hot-electron temperature turned out to be faster than that estimated in typical pump-probe optical experiments~(see e.g.~Ref.~\citenum{Ferrera_ACSphotonics_2020} and references therein). This is well captured by the $G$ parameter of the 3TM whose value obtained from the simulations turned out to be about 3 times smaller than the standard value assumed for 3TM analysis. Such a discrepancy arises from the fact that, differently to optical pump-probe measurements, our experiments are performed at low temperature (137 K), and it is well known that the $G$ coefficient decreases with the lattice temperature. However, the exact temperature dependence of $G$ is still debated, with experimental results (e.g.~\cite{Mueller_ASS_2014}) that largely overestimate theoretical calculations (e.g.~\cite{Medvedev_PRB_2020}). Our results indicate that the value of $G$ at low temperatures is definitely much lower than that reported in previous measurements, and about 60\% larger than in the atomistic calculations by Medvedev and Milov~\cite{Medvedev_PRB_2020}.



\newpage
\section{Conclusion}

We performed ultrafast pump-probe photoemission experiments across the Fermi-edge region of photoexcited Au NPs, and observed a subtle, yet clear variation of the PE spectra as a function of the time $\tau$ after the excitation,
ascribed to the ultrafast dynamics of the electron-gas temperature $\vartheta_e$.
The data showed a fast rise of $\vartheta_e$ within the first few hundred femtoseconds, followed by a slow decay in the first picoseconds, in agreement with the general picture of ultrafast electrodynamics in metallic NPs.
The $\vartheta_e$ measurement relied on the definition itself of electronic temperature via the Fermi-Dirac distribution function.
In this sense, the method can be exploited to extract the temperature dynamics devoid of any assumption of intermediate modelling step,
paving the way for the assessment of the relaxation dynamics within physical systems that are, at present, too complex to handle theoretically, (at least with atomistic calculations). Also, our results points in favor of a review of some key parameters commonly employed in the thermodynamic Three-Temperature Model, and support recent findings from atomistic calculations of electron-electron and electron-phonon scattering rates.

Experiments like the one reported here are still technically challenging, as multiple stringent requirements have to be simultaneously fulfilled,
but future developments in instrumentation and experimental design promise exciting applications of the method.
In this respect, we believe that this work will represent a reference point for experimental and theoretical
investigations of the ultrafast dynamics 
and open the way for direct and quantitatively accurate studies of the electronic properties of metallic nanosystems.

\newpage

\section{Experimental Section}

\subsubsection{Preparation of AZO film}The film was deposited on MgO(001) substrate via magnetron sputtering by co-deposition from
a 3-in. RF magnetron source (ZnO) and a 3-in. DC magnetron
source (Al) operating in confocal geometry approximately 15 cm from the substrate.\cite{Benedetti2015} 
A constant 0.7 {\AA}/s ZnO deposition rate was reached at RF power of 120 W. 
The DC power was varied in order to obtain a doping level
of 4-at. \%. 
The deposition was performed at $300^{\circ}$C with base pressure of
1$\cdot 10^{-6}$ mbar, in a 5$\cdot 10^{-3}$ mbar Ar atmosphere. 
The morphopolgy of the surface of the AZO film was examined by atomic force microscopy (AFM), as detailed in Figure S1 of SI. The resistivity of the film was around $\rho\simeq 10^{-4} \Omega$$\cdot$cm.\\
\subsubsection{Deposition of Au NPs}The Au NPs were deposited onto a  thick ($\approx$ 150 nm) Al-doped ZnO (AZO) film. Au NPs were obtained by solid-state dewetting (at $400^{\circ}$C) of a 3-nm-thick Au film grown by molecular-beam epitaxy at $\sim 10^{-9}$ mbar pressure. Au was deposited at room temperature by evaporation from a Mo crucible, at $60^{\circ}$ of incidence with respect to the surface normal. Mild annealing at $400^{\circ}$C induced the dewetting of Au and the formation of closely disconnected NPs.
The Au NPs had an areal density of about $600 \pm 100$~NP$/\mu$m$^2$,
a mean size of around 20-25 nm in diameter, circular in-plane cross-section (see the atomic-force microscope image in Figure S2), and they exhibited a room-temperature LSP at $\lambda_{R}\approx600$ nm.\\
\subsubsection{Tr-PES Pump-Probe Setup}The pump was an ultrashort pulse with wavelength $\lambda= 650$ nm (spot size $\approx1$~mm (FWHM), maximum energy per pulse $\approx0.5$~$\mu$J, duration $\approx100$~\textit{fs}) whereas the probe was an EUV pulse obtained by HHG (photon energy 16.9 eV, linearly polarized, spot size $\approx200$~$\mu$m (FWHM), energy resolution $\approx30$~meV, time duration $\approx100$~\textit{fs}).
The pump and probe beams are recombined, with a variable delay, in the beamline endstation, at the sample location (see the scheme in Figure \ref{trans_abs}). 
The spatial overlap between pump and probe beams was verified by superimposing the two beams on a cerium-doped YAG scintillator.
The end station is an ultra-high vacuum chamber equipped with a Scienta electron analyser (200 mm radius, maximum energy resolution 2 meV). 
The angle of incidence of the pump beam was $\theta=65^{\circ}$ and the beam was linearly polarized orthogonal to the incidence plane
({\em s}-polarization).
The sample holder was cooled by liquid nitrogen flow.
The energy resolution of the PE setup, determined from independent measurements \cite{Cucini2020}, was equal to $w=35$ meV. 
The zero delay has been measured using the sum-frequency generation (SFG) signal in a BBO crystal, overlapping the pump and the HHG laser seed (515 nm, 2.4 eV), which is intrinsically time overlapped with the EUV harmonic. 
Since SFG is used to measure the pulse duration, the time-delay uncertainty is smaller than the pulse duration.
Above $\tau=1$~ps we collected only sparse points, since the main focus of our investigation lies in the
short-delay regime, where $\vartheta_e$ exhibits the most interesting behaviour, and since other experimental methods
are available for {\em directly} assessing the system temperature after electrons and lattice have thermalized \cite{Ferrera_ACSphotonics_2020}. 
The pump wavelength is spectrally overlapped with the tail of the LSP resonance, ensuring a sizable excitation cross section of the NPs.
The applicable pump fluence was however limited  due to the onset of
pump-induced nonlinear photoemission from the system that would smear the PE spectra due to space-charge effects \cite{spacechargeeffects_DC1,spacechargeeffects_DC2,spacechargeeffects_DC3}.
A fluence of $F= 0.19\;\mathrm{J/m^2}$ (peak intensity of $1.86 \times10^{12}$~W/m$^2$) was chosen as the best compromise between weak-enough space-charge effects
and reasonable energy deposition into the NPs.
The overall photoelectron yield was kept low for analogous space-charge reasons.
The limited fluence implies that only moderate increases of the electron temperature $\vartheta_e$ can be achieved.
In order to observe such relatively weak perturbations of the Fermi-Dirac (FD) distribution, both a high signal-to-noise ratio and a high spectral resolution are needed, although the simultaneous fulfilment of both requirements is intrinsically challenging.
\subsubsection{Three-temperature model} The 3TM is a thermodynamic model describing the energy flow dynamics following photoexcitation of noble metal structures in terms of the three energetic degrees of freedom: the excess energy density of nonthermal electrons ($N$), the temperature of thermal electrons ($\theta_e$), and the lattice temperature ($\theta_l$). This dynamics is detailed by a system of rate equations reading\cite{Sun_PRB_1994}:
\begin{eqnarray}
\frac{d N}{d t}&=& P_a(t) - N/\tau_{ee} \label{ettm1}\\
\gamma \theta_e\frac{d \theta_e}{d t}&=& N/\tau_{ee} -G(\theta_e - \theta_l) \label{ettm2}\\
C_l \frac{d \theta_l}{d t}&=& G(\theta_e - \theta_l) \label{ettm3}
\end{eqnarray}
\noindent where $\gamma = 68$~J~m$^{-3}$~K$^{-2}$ is the electrons heat capacity constant, $C_l = 2.5 \times 10^6$~J~m$^{-3}$~K$^{-1}$ is the lattice heat capacity, $G$ is the electron-phonon (electron-ion) coupling constant, and $\tau_{ee}$ is the electron-electron thermalization time constant (see main text for the values of $G$ and $\tau_{ee}$ used in the calculations). In Equations (\ref{ettm1})-(\ref{ettm3}), $P_a(t)$ is the absorbed power density, computed from the effective absorption cross-section of the nanostructure at the pump wavelength, $\sigma_{A}^{eff}(\lambda_P)$, according to the formula:
\begin{equation}\label{Pabs}
P_a(t) = \frac{\sigma_{A}^{eff}(\lambda_P)F}{V \tau_P} g(t),
\end{equation}
where $F$ is the incident fluence (defined as the ratio between the pump pulse energy and the effective area of the pump beam), $\tau_P = 120$~fs is the duration of the pump pulse, FWHM (50\% larger than the measured duration to take into account convolutional effects with the probe pulse), and $V$ is the (average) volume of the nanoparticle. In above equation, $g(t)$ is the normalized Gaussian function describing the temporal profile of the pulse: 
\begin{equation}\label{gdt}
g(t) =\sqrt{\frac{4 \ln(2)}{\pi}} \exp \left(-\frac{4\ln(2)t^2}{\tau_P^2} \right).
\end{equation}
In our simulations $\sigma_{A}^{eff}(\lambda_P) = 225$~nm$^2$, estimated from quasi-static formulas.

\medskip
\textbf{Supporting Information} \par
Supporting Information is available from the Wiley Online Library or from the author.

\medskip
\textbf{Acknowledgements} \par

\begin{wrapfigure}{l}{0.1\textwidth}
  \vspace{-30pt}
  \begin{center}
    \includegraphics[width=0.1\textwidth]{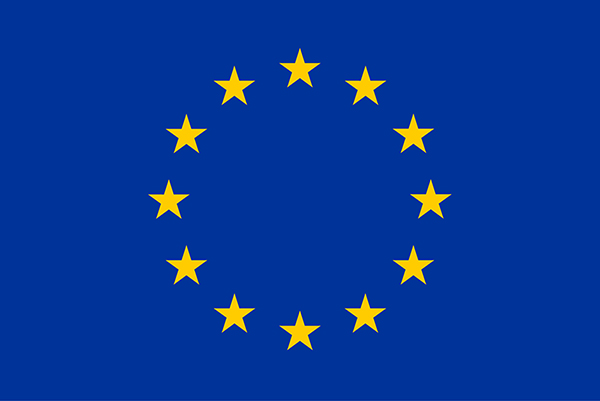}
  \end{center}
  \vspace{-20pt}
  \vspace{-10pt}
\end{wrapfigure}

This project has received funding from the European Union's Horizon 2020 research and innovation programme under the Marie Sk\l odowska-Curie grant agreement N$^{\circ}$799126. \\This work has been performed in the framework of the Nanoscience Foundry and Fine Analysis (NFFA-MIUR Italy Progetti Internazionali) facility.


\begin{thebibliography}{10}
\providecommand{\url}[1]{\texttt{#1}}
\providecommand{\urlprefix}{URL }

\bibitem{abinitio2016}
A.~M. Brown, R.~Sundararaman, P.~Narang, W.~A. Goddard, H.~A. Atwater,
\newblock \emph{Phys. Rev. B} \textbf{2016}, \emph{94} 075120.

\bibitem{Bisio2017}
F.~Bisio, E.~Principi, M.~Magnozzi, A.~Simoncig, E.~Giangrisostomi,
  R.~Mincigrucci, L.~Pasquali, C.~Masciovecchio, F.~Boscherini, M.~Canepa,
\newblock \emph{Phys. Rev. B} \textbf{2017}, \emph{96} 081119.

\bibitem{Saavedra2016}
J.~R.~M. Saavedra, A.~Asenjo-Garcia, F.~J. García~de Abajo,
\newblock \emph{ACS Photonics} \textbf{2016}, \emph{3}, 9 1637.

\bibitem{Yonatan2019}
Y.~Dubi, Y.~Sivan,
\newblock \emph{Light Sci. Appl.} \textbf{2019}, \emph{8}, 1 89.

\bibitem{Ferrera_ACSphotonics_2020}
M.~Ferrera, G.~Della~Valle, M.~Sygletou, M.~Magnozzi, D.~Catone, P.~O’Keeffe,
  A.~Paladini, F.~Toschi, L.~Mattera, M.~Canepa, F.~Bisio,
\newblock \emph{ACS Photonics} \textbf{2020}, \emph{7}, 4 959.

\bibitem{Kumar2016}
S.~Kumar, A.~K. Sood,
\newblock \emph{Ultrafast Response of Plasmonic Nanostructures}, 131--167,
\newblock Springer International Publishing, \textbf{2016}.

\bibitem{Brongersma2015}
M.~L. Brongersma, N.~J. Halas, P.~Nordlander,
\newblock \emph{Nat. Nanotechnol.} \textbf{2015}, \emph{10}, 1 25.

\bibitem{Baffou2011}
G.~Baffou, H.~Rigneault,
\newblock \emph{Phys. Rev. B} \textbf{2011}, \emph{84} 035415.

\bibitem{Guo2013}
S.~Guo, S.~Zhang, S.~Sun,
\newblock \emph{Angew. Chem} \textbf{2013}, \emph{52}, 33 8526.

\bibitem{bian2013}
Z.~Bian, T.~Tachikawa, P.~Zhang, M.~Fujitsuka, T.~Majima,
\newblock \emph{J. Am. Chem. Soc.} \textbf{2014}, \emph{136}, 1 458.

\bibitem{Clavero2014b}
C.~Clavero,
\newblock \emph{Nat. Photonics} \textbf{2014}, \emph{8} 95.

\bibitem{kale2014}
M.~J. Kale, T.~Avanesian, P.~Christopher,
\newblock \emph{ACS Catal} \textbf{2014}, \emph{4}, 1 116.

\bibitem{Ferry2010}
V.~E. Ferry, J.~N. Munday, H.~A. Atwater,
\newblock \emph{Adv. Mater.} \textbf{2010}, \emph{22}, 43 4794.

\bibitem{Atwater2010}
H.~A. Atwater, A.~Polman,
\newblock \emph{Nat. Mater.} \textbf{2010}, \emph{9}, 3 205.

\bibitem{kovalenko2010}
D.~V. Talapin, J.-S. Lee, M.~V. Kovalenko, E.~V. Shevchenko,
\newblock \emph{Chem. Rev.} \textbf{2010}, \emph{110}, 1 389.

\bibitem{Stratakis2013}
E.~Stratakis, E.~Kymakis,
\newblock \emph{Mater. Today} \textbf{2013}, \emph{16}, 4 133 .

\bibitem{Sygletou2017}
M.~Sygletou, C.~Petridis, E.~Kymakis, E.~Stratakis,
\newblock \emph{Adv. Mater.} \textbf{2017}, \emph{29}, 39 1700335.

\bibitem{liu2018}
J.~G. Liu, H.~Zhang, S.~Link, P.~Nordlander,
\newblock \emph{ACS Photonics} \textbf{2018}, \emph{5}, 7 2584.

\bibitem{smith2018}
K.~J. Smith, Y.~Cheng, E.~S. Arinze, N.~E. Kim, A.~E. Bragg, S.~M. Thon,
\newblock \emph{ACS Photonics} \textbf{2018}, \emph{5}, 3 805.

\bibitem{Kravets2018}
V.~G. Kravets, A.~V. Kabashin, W.~L. Barnes, A.~N. Grigorenko,
\newblock \emph{Chem. Rev.} \textbf{2018}, \emph{118}, 12 5912.

\bibitem{SEPULVEDA2009}
B.~Sep\'ulveda, P.~C. Angelom\'e, L.~M. Lechuga, L.~M. Liz-Marz\'an,
\newblock \emph{Nano Today} \textbf{2009}, \emph{4}, 3 244 .

\bibitem{Manjavacas2014}
A.~Manjavacas, J.~G. Liu, V.~Kulkarni, P.~Nordlander,
\newblock \emph{ACS Nano} \textbf{2014}, \emph{8}, 8 7630.

\bibitem{Taylor2014}
A.~B. Taylor, A.~M. Siddiquee, J.~W.~M. Chon,
\newblock \emph{ACS Nano} \textbf{2014}, \emph{8}, 12 12071.

\bibitem{Herrmann2014}
L.~O. Herrmann, V.~K. Valev, C.~Tserkezis, J.~S. Barnard, S.~Kasera, O.~A.
  Scherman, J.~Aizpurua, J.~J. Baumberg,
\newblock \emph{Nat. Commun.} \textbf{2014}, \emph{5} 4568.

\bibitem{Gonzalez2017}
G.~Gonz{\'a}lez-Rubio, P.~D{\'\i}az-N{\'u}{\~n}ez, A.~Rivera, A.~Prada,
  G.~Tardajos, J.~Gonz{\'a}lez-Izquierdo, L.~Ba{\~n}ares, P.~Llombart, L.~G.
  Macdowell, M.~Alcolea~Palafox, L.~M. Liz-Marz{\'a}n,
  O.~Pe{\~n}a-Rodr{\'\i}guez, A.~Guerrero-Mart{\'\i}nez,
\newblock \emph{Science} \textbf{2017}, \emph{358}, 6363 640.

\bibitem{Catone2018}
D.~Catone, A.~Ciavardini, L.~Di~Mario, A.~Paladini, F.~Toschi, A.~Cartoni,
  I.~Fratoddi, I.~Venditti, A.~Alabastri, R.~Proietti~Zaccaria, P.~O{'}Keeffe,
\newblock \emph{J. Phys. Chem. Lett.} \textbf{2018}, \emph{9}, 17 5002.

\bibitem{Christofilos2003}
A.~Arbouet, C.~Voisin, D.~Christofilos, P.~Langot, N.~D. Fatti, F.~Vall\'ee,
  J.~Lerm\'e, G.~Celep, E.~Cottancin, M.~Gaudry, M.~Pellarin, M.~Broyer,
  M.~Maillard, M.~P. Pileni, M.~Treguer,
\newblock \emph{Phys. Rev. Lett.} \textbf{2003}, \emph{90} 177401.

\bibitem{loukakos2004}
C.~Voisin, D.~Christofilos, P.~A. Loukakos, N.~Del~Fatti, F.~Vall\'ee,
  J.~Lerm\'e, M.~Gaudry, E.~Cottancin, M.~Pellarin, M.~Broyer,
\newblock \emph{Phys. Rev. B} \textbf{2004}, \emph{69} 195416.

\bibitem{Minutella2017}
E.~Minutella, F.~Schulz, H.~Lange,
\newblock \emph{J. Phys. Chem. Lett.} \textbf{2017}, \emph{8}, 19 4925.

\bibitem{Hartland2017}
G.~V. Hartland, L.~V. Besteiro, P.~Johns, A.~O. Govorov,
\newblock \emph{ACS Energy Lett.} \textbf{2017}, \emph{2}, 7 1641.

\bibitem{Zhang2018}
X.~Zhang, C.~Huang, M.~Wang, P.~Huang, X.~He, Z.~Wei,
\newblock \emph{Sci. Rep.} \textbf{2018}, \emph{8} 10499.

\bibitem{gallardo2019}
O.~A. Douglas-Gallardo, M.~Berdakin, T.~Frauenheim, C.~G. Sánchez,
\newblock \emph{Nanoscale} \textbf{2019}, \emph{11} 8604.

\bibitem{Stoll2014}
T.~Stoll, P.~Maioli, A.~Crut, N.~Fatti, F.~Vallée,
\newblock \emph{Eur. Phys. J. B} \textbf{2014}, \emph{87} 260.

\bibitem{Magnozzi2019}
M.~Magnozzi, M.~Ferrera, L.~Mattera, M.~Canepa, F.~Bisio,
\newblock \emph{Nanoscale} \textbf{2019}, \emph{11} 1140.

\bibitem{Ferrera2019}
M.~Ferrera, M.~Magnozzi, F.~Bisio, M.~Canepa,
\newblock \emph{Phys. Rev. Mater.} \textbf{2019}, \emph{3} 105201.

\bibitem{Sayed2001}
M.~B. Mohamed, T.~S. Ahmadi, S.~Link, M.~Braun, M.~A. El-Sayed,
\newblock \emph{Chem. Phys. Lett.} \textbf{2001}, \emph{343}, 1 55 .

\bibitem{Feldstein1997}
M.~J. Feldstein, C.~D. Keating, Y.-H. Liau, M.~J. Natan, N.~F. Scherer,
\newblock \emph{J. Am. Chem. Soc.} \textbf{1997}, \emph{119}, 28 6638.

\bibitem{Hartland1998}
J.~H. Hodak, I.~B. Martini, G.~V. Hartland,
\newblock \emph{Chem. Phys. Lett.} \textbf{1998}, \emph{284} 135.

\bibitem{Wang2015}
X.~Wang, Y.~Guillet, P.~R. Selvakannan, H.~Remita, B.~Palpant,
\newblock \emph{J. Phys. Chem. C} \textbf{2015}, \emph{119}, 13 7416.

\bibitem{Harutyunyan2015}
H.~Harutyunyan, A.~B.~F. Martinson, D.~Rosenmann, L.~K. Khorashad, L.~V.
  Besteiro, A.~O. Govorov, G.~P. Wiederrecht,
\newblock \emph{Nat. Nanotechnol.} \textbf{2015}, \emph{10}, 9 770.

\bibitem{Broek2011}
B.~Van~de Broek, D.~Grandjean, J.~Trekker, J.~Ye, K.~Verstreken, G.~Maes,
  G.~Borghs, S.~Nikitenko, L.~Lagae, C.~Bartic, K.~Temst, M.~J. Van~Bael,
\newblock \emph{Small} \textbf{2011}, \emph{7}, 17 2498.

\bibitem{Plech2017}
A.~Plech, S.~Ibrahimkutty, S.~Reich, G.~Newby,
\newblock \emph{Nanoscale} \textbf{2017}, \emph{9} 17284.

\bibitem{Richardson2009}
H.~H. Richardson, M.~T. Carlson, P.~J. Tandler, P.~Hernandez, A.~O. Govorov,
\newblock \emph{Nano Letters} \textbf{2009}, \emph{9}, 3 1139.

\bibitem{Petek2021}
A.~Li, M.~Reutzel, Z.~Wang, D.~Novko, B.~Gumhalter, H.~Petek,
\newblock \emph{ACS Photonics} \textbf{2021}, \emph{8}, 1 247.

\bibitem{Cucini2020}
R.~Cucini, T.~Pincelli, G.~Panaccione, D.~Kopic, F.~Frassetto, P.~Miotti, G.~M.
  Pierantozzi, S.~Peli, A.~Fondacaro, A.~De~Luisa, A.~De~Vita, P.~Carrara,
  D.~Krizmancic, D.~T. Payne, F.~Salvador, A.~Sterzi, L.~Poletto,
  F.~Parmigiani, G.~Rossi, F.~Cilento,
\newblock \emph{Struct. Dyn.} \textbf{2020}, \emph{7}, 1 014303.

\bibitem{smearing2004}
H.~H\"ovel, I.~Barke, H.-G. Boyen, P.~Ziemann, M.~G. Garnier, P.~Oelhafen,
\newblock \emph{Phys. Rev. B} \textbf{2004}, \emph{70} 045424.

\bibitem{Kurtz2007}
R.~L. Kurtz, D.~A. Browne, G.~J. Mankey,
\newblock \emph{J. Phys. Condens. Matter} \textbf{2007}, \emph{19}, 35 355001.

\bibitem{theoretical2014}
R.~Sundararaman, P.~Narang, A.~S. Jermyn, W.~A. Goddard~III, H.~A. Atwater,
\newblock \emph{Nat. Commun.} \textbf{2014}, \emph{5}, 1 5788.

\bibitem{theory2015}
M.~Bernardi, J.~Mustafa, J.~B. Neaton, S.~G. Louie,
\newblock \emph{Nat. Commun.} \textbf{2015}, \emph{6}, 1 7044.

\bibitem{ACSNANO2020}
T.~P. Rossi, P.~Erhart, M.~Kuisma,
\newblock \emph{ACS Nano} \textbf{2020}, \emph{14}, 8 9963.

\bibitem{Govorov2013}
A.~O. Govorov, H.~Zhang, Y.~K. Gun’ko,
\newblock \emph{The Journal of Physical Chemistry C} \textbf{2013}, \emph{117},
  32 16616.

\bibitem{Govorov2014}
A.~O. Govorov, H.~Zhang, H.~V. Demir, Y.~K. Gun’ko,
\newblock \emph{Nano Today} \textbf{2014}, \emph{9}, 1 85 .

\bibitem{Chang_ACSEL_2019}
L.~Chang, L.~V. Besteiro, J.~Sun, E.~Y. Santiago, S.~K. Gray, Z.~Wang, A.~O.
  Govorov,
\newblock \emph{ACS Energy Letters} \textbf{2019}, \emph{4}, 10 2552.

\bibitem{tzortzakis2000}
N.~Del~Fatti, C.~Voisin, M.~Achermann, S.~Tzortzakis, D.~Christofilos,
  F.~Vall\'ee,
\newblock \emph{Phys. Rev. B} \textbf{2000}, \emph{61} 16956.

\bibitem{hartland2004}
G.~V. Hartland,
\newblock \emph{Phys. Chem. Chem. Phys.} \textbf{2004}, \emph{6} 5263.

\bibitem{groeneveld1992}
R.~H.~M. Groeneveld, R.~Sprik, A.~Lagendijk,
\newblock \emph{Phys. Rev. B} \textbf{1992}, \emph{45} 5079.

\bibitem{sun1994}
C.-K. Sun, F.~Vall\'ee, L.~H. Acioli, E.~P. Ippen, J.~G. Fujimoto,
\newblock \emph{Phys. Rev. B} \textbf{1994}, \emph{50} 15337.

\bibitem{Imamura2006}
M.~Imamura, A.~Tanaka,
\newblock \emph{Phys. Rev. B} \textbf{2006}, \emph{73}.

\bibitem{Sun_PRB_1994}
C.-K. Sun, F.~Vall{\'e}e, L.~Acioli, E.~Ippen, J.~Fujimoto,
\newblock \emph{Phys. Rev. B} \textbf{1994}, \emph{50}, 20 15337.

\bibitem{Zavelani_ACSP_2015}
M.~Zavelani-Rossi, D.~Polli, S.~Kochtcheev, A.-L. Baudrion, J.~B\'eal,
  V.~Kumar, E.~Molotokaite, M.~Marangoni, S.~Longhi, G.~Cerullo, P.-M. Adam,
  G.~D. Valle,
\newblock \emph{ACS Photonics} \textbf{2015}, \emph{2}, 4 521.

\bibitem{Della_Valle_JPCL_2013}
G.~Della~Valle, F.~Scotognella, A.~R. Srimath~Kandada, M.~Zavelani-Rossi,
  H.~Li, M.~Conforti, S.~Longhi, L.~Manna, G.~Lanzani, F.~Tassone,
\newblock \emph{J. Phys. Chem. Lett.} \textbf{2013}, \emph{4} 3337.

\bibitem{Gaspari_NL_2017}
R.~Gaspari, G.~Della~Valle, S.~Ghosh, I.~Kriegel, F.~Scotognella, A.~Cavalli,
  L.~Manna,
\newblock \emph{Nano Lett.} \textbf{2017}, \emph{17}, 12 7691.

\bibitem{Della_Valle_PRB_2015}
G.~Della~Valle, D.~Polli, P.~Biagioni, C.~Martella, M.~C. Giordano, M.~Finazzi,
  S.~Longhi, L.~Du\`o, G.~Cerullo, F.~Buatier~de Mongeot,
\newblock \emph{Phys. Rev. B} \textbf{2015}, \emph{91} 235440.

\bibitem{Silva_PRB_2018}
M.~G. Silva, D.~C. Teles-Ferreira, L.~Siman, C.~R. Chaves, L.~O. Ladeira,
  S.~Longhi, G.~Cerullo, C.~Manzoni, A.~M. de~Paula, G.~Della~Valle,
\newblock \emph{Phys. Rev. B} \textbf{2018}, \emph{98} 115407.

\bibitem{Bouillard_NL_2012}
J.-S.~G. Bouillard, W.~Dickson, D.~P. O’Connor, G.~A. Wurtz, A.~V. Zayats,
\newblock \emph{Nano Letters} \textbf{2012}, \emph{12}, 3 1561, pMID: 22339644.

\bibitem{Bernardi_NatComm_2015}
M.~Bernardi, J.~Mustafa, J.~Neaton, S.~G. Louie,
\newblock \emph{Nat. Commun.} \textbf{2015}, \emph{6} 7044.

\bibitem{Schirato_NatComm_2020}
A.~Schirato, M.~Maiuri, A.~Toma, S.~Fugattini, R.~{Proietti Zaccaria},
  P.~Laporta, P.~Nordlander, G.~Cerullo, A.~Alabastri, G.~{Della Valle},
\newblock \emph{Nat. Photon.} \textbf{2020}, \emph{14} 723.

\bibitem{Mueller_ASS_2014}
B.~Mueller, B.~Rethfeld,
\newblock \emph{Appl. Surf. Sci.} \textbf{2014}, \emph{302} 24.

\bibitem{Medvedev_PRB_2020}
N.~Medvedev, I.~Milov,
\newblock \emph{Phys. Rev. B} \textbf{2020}, \emph{102} 064302.

\bibitem{Benedetti2015}
I.~Valenti, S.~Benedetti, A.~di~Bona, V.~Lollobrigida, A.~Perucchi,
  P.~Di~Pietro, S.~Lupi, S.~Valeri, P.~Torelli,
\newblock \emph{Journal of Applied Physics} \textbf{2015}, \emph{118}, 16
  165304.

\bibitem{spacechargeeffects_DC1}
L.~Oloff, M.~Oura, K.~Rossnagel, A.~Chainani, M.~Matsunami, R.~Eguchi, T.~Kiss,
  Y.~Nakatani, T.~Yamaguchi, J.~Miyawaki, M.~Taguchi, K.~Yamagami, T.~Togashi,
  T.~Katayama, K.~Ogawa, M.~Yabashi, T.~Ishikawa,
\newblock \emph{New J. Phys.} \textbf{2014}, \emph{16} 123045.

\bibitem{spacechargeeffects_DC2}
L.~Oloff, A.~Chainani, M.~Matsunami, K.~Takahashi, T.~Togashi, H.~Osawa,
  K.~Hanff, A.~Quer, R.~Matsushita, R.~Shiraishi, M.~Nagashima, A.~Kimura,
  K.~Matsuishi, M.~Yabashi, Y.~Tanaka, G.~Rossi, T.~Ishikawa, K.~Rossnagel,
  M.~Oura,
\newblock \emph{Sci. Rep.} \textbf{2016}, \emph{6} 35087.

\bibitem{spacechargeeffects_DC3}
D.~Kühn, E.~Giangrisostomi, R.~Jay, N.~Sorgenfrei, A.~Foehlisch,
\newblock \emph{New J. Phys.} \textbf{2019}, \emph{21} 073042.

\end{thebibliography}

\newpage




\begin{figure}
\centering
\medskip
\includegraphics[width=0.75\linewidth]{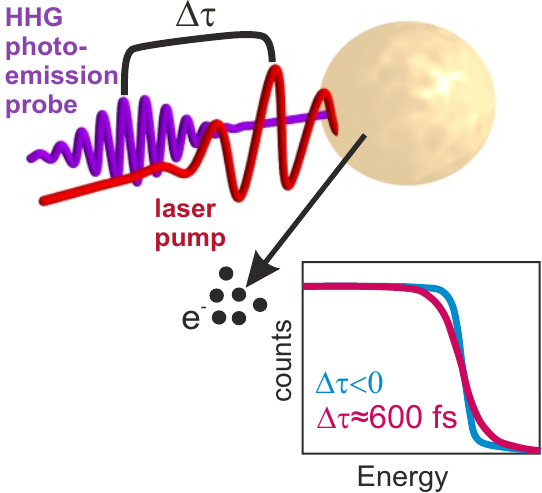}
  \medskip
  \caption*{ToC Entry}
\end{figure}

\end{document}